\documentclass{mem}
\usepackage{natbib}
\usepackage{txfonts}
\usepackage{balance}
\usepackage{graphicx}
\usepackage[a4paper]{hyperref}
\usepackage{txfonts}
\usepackage{natbib}
\usepackage{amssymb}
\bibpunct{(}{)}{;}{a}{}{,} 
\idline{75}{282}
\def\cres{CR{\itshape e}s}
\def\crps{CR{\itshape p}s}
\begin{document}

\title{
Non-thermal emission from galaxy clusters
}

   \subtitle{}

\author{
C. Ferrari\inst{1} 
          }

  \offprints{C. Ferrari}

\institute{
UNS, CNRS, OCA, Laboratoire Cassiop\'ee, Nice, France\\
\email{chiara.ferrari@oca.eu}}

\authorrunning{Ferrari }

\titlerunning{Non-thermal emission from galaxy clusters}

\abstract{The relevance of non-thermal cluster studies and the importance of combining observations of future radio surveys with {\it WFXT} data are discussed in this paper.
}
\maketitle{}

\section{Introduction}

The discovery of diffuse radio sources in a few tens of merging galaxy
clusters has pointed out the existence of a non-thermal component
(i.e. relativistic particles with Lorentz factor $\gamma>>$1000 and
magnetic fields of the order of $\mu$G) in the intracluster volume
\citep[e.g][]{2008SSRv..134...93F}.  Through non-thermal studies of
galaxy clusters we can estimate the cosmic-ray and magnetic field
energy budget and pressure contribution to the intracluster medium
(ICM), as well as get clues about the cluster dynamical state and
energy redistribution during merging events. Non-thermal analyses can
elucidate non-equilibrium physical processes whose deep understanding
is essential to do high-precision cosmology using galaxy clusters
\citep{2008MNRAS.385.1242P}.

In the following, we will give an overview of the main open questions
about the non-thermal intracluster component (Sect.~\ref{sec:open}).
The perspectives that will be opened in this field by a new generation
of radio telescopes will also be addressed (Sect.~\ref{sec:radio}). We
will focus in particular on the study of clusters with similar X-ray
and radio morphologies, i.e. clusters hosting diffuse radio sources
that are called ``radio halos''. The importance of an X-ray facility
such as {\it WFXT} will be discussed (Sect.~\ref{sec:wfxt}). The
$\Lambda$CDM model with H$_0$=70 km ${\rm s}^{-1} {\rm Mpc}^{-1}$,
$\Omega_m=0.3$ and $\Omega_{\Lambda}=0.7$ has been adopted.

\section{Open questions} \label{sec:open}

The origin of intracluster cosmic rays (CRs) is matter of debate. CRs,
gyrating around magnetic field lines which are frozen in the ICM, have
typical diffusion velocity of the order of the Alfv\'en speed ($\sim$
100 km/s). They thus need $\gtrsim$10 Gyr to propagate over radio halo
extensions.  Radiative timescales are longer than the Hubble time for
CR protons (\crps). They thus can be continuously accelerated
(directly in the ICM or inside active galaxies and then ejected),
resulting in an effective accumulation of relativistic and
ultra-relativistic \crps~in clusters. Hadronic CRs can subsequently
produce Gamma-rays and secondary relativistic electrons through
inelastic collisions with the ions of the ICM
\citep[e.g.][]{2009A&A...495...27A}.

The radiative lifetime of relativistic electrons (\cres) is instead
much shorter ($\lesssim$0.1 Gyr) than their cluster crossing time due
to inverse Compton (IC) and synchrotron energy losses. Therefore
\cres~ have to be continuously re-accelerated {\it in situ}.  Two main
classes of models have been proposed to explain intracluster electron
acceleration: primary and secondary models. The former predict the
acceleration of fossil radio plasma or directly of thermal electrons
of the ICM through shocks and/or MHD turbulence generated by cluster
mergers. Secondary models predict instead that non-thermal electrons
in clusters are the secondary product of hadronic interactions between
relativistic protons and the ions of the thermal ICM
\citep[e.g.][]{2008SSRv..134...93F}.

Current observational results are in favour of primary models. The
very few detailed analyses of the radio spectral index
$\alpha$\footnote{$S_{\nu} \propto \nu^{-\alpha}$} distribution in
radio halos show hints of a possible increase of $\alpha$ as a
function of radius and of frequency
\citep[e.g][]{2003A&A...397...53T}, as expected in the case of primary
models. A possible anti-correlation between $\alpha$ and the ICM
temperature (i.e. flatter spectra in hotter regions) has also been
pointed out in a few cases \citep[e.g.][]{2007A&A...467..943O}. The
hottest ICM regions are usually associated to shock and/or turbulence
induced by cluster collisions. The fact that these regions host
younger \cres~ is thus in agreement with primary models. A unique
prediction of the turbulence re-acceleration models is the existence
of ultra-steep radio halos, not associated to major cluster mergers,
but to less energetic merging events. Recently
\citet{2008Natur.455..944B} claimed the detection of the first
ultra-steep radio halo in the multiple merging cluster A521. At the
moment, the most striking observational evidence in favour of primary
models is the fact that diffuse cluster sources have been detected
only in merging clusters.

Deeper statistical analyses of the correlation between diffuse radio
sources and the physical properties of their host clusters are
required to refine the physical models for CR acceleration. For
instance, we know that diffuse radio sources have been detected in
$\leq$ 10\% of known clusters, while about 40\% of clusters show a
disturbed dynamical state: why cluster mergers seem to be a necessary
but not sufficient condition for the acceleration of intracluster
relativistic particles? The answer could be related to the cluster
mass, since a correlation between radio and X-ray cluster luminosity
has been pointed out
\citep[e.g.][]{2001ApJ...553L..15B}. This suggests that
only the most massive merging clusters are energetic enough to produce
diffuse radio emission at power levels observable with current radio
observations (see also the discussion in Sect. \ref{sec:radio}).

Even more debated are the origin and properties of intracluster
magnetic fields \citep{2008SSRv..134..311D}. The different methods
available to measure intracluster magnetic fields (equipartition
assumption, rotation measures, Compton scattering of CMB photons,
X-ray study of cooling-cores in the ICM) show quite discrepant results
(see Table 3 of \citet{2004IJMPD..13.1549G}). Different reasons can
explain this discrepancy \citep[e.g.][]{2010arXiv1005.3699F}. Again,
higher statistics is required for magnetic field measurements. For
instance, we need deeper multi-wavelength radio observations of radio
galaxies per cluster for rotation measure (RM) estimates, combined to
detailed modeling of the ICM X-ray brightness profile
\citep{2001A&A...379..807G}.

\section{Perspectives} \label{sec:est}

\subsection{A new generation of radio telescopes} \label{sec:radio}

\begin{figure}
\resizebox{\hsize}{!}{
\includegraphics{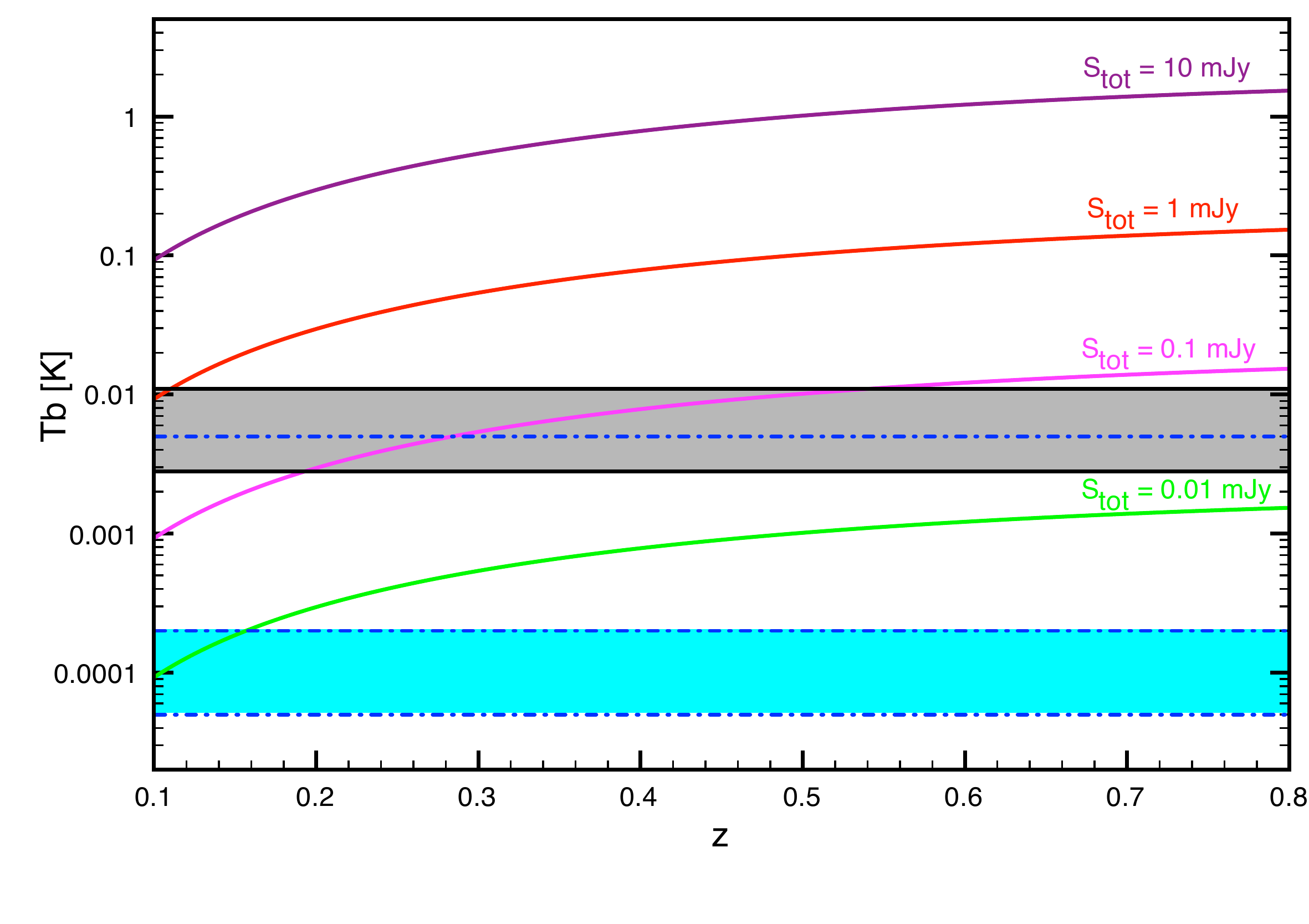}}
\caption{\footnotesize Brightness temperature at $\sim$ 1.3 GHz as a
    function of redshift expected for radio halos of a given total
    flux and of 0.5 Mpc radius. Possible limits for the {\it ASKAP} EMU
    Survey and for 50\% of the {\it SKA} with 1 hour integration time are
    indicated (shaded areas delimited by solid and dot-dashed lines
    respectively).}
\label{fig:fig1}
\end{figure}

Huge perspectives in the study of the non-thermal intracluster
component will be open by the Low Frequency Array \citep[{\it
  LOFAR},][]{2006astro.ph.10596R}. A steepening of the synchrotron
spectrum of radio halos is expected in the framework of stochastic
particle acceleration by MHD turbulence and it has been observed in
several halos
\cite[see][]{2008SSRv..134...93F}. \citet{2010A&A...509A..68C} have
introduced a characteristic frequency $\nu_s \sim 7\nu_b$ at which the
steepening become extremely severe. Basically, surveys at frequency
$\nu > \nu_s$ cannot detect radio halos. Since the lower is the radio
luminosity of the halo, the lower is the expected break frequency,
high-frequency ($\nu \approx$ 1.4 GHz) surveys are sensitive only to
high-luminosity halos, while most of the faint radio luminosity halo
tail will appear at low-frequency radio observations. That's the huge
potential of {\it LOFAR} in the study of diffuse cluster radio
sources. Radio maps resulting from the {\it LOFAR}
Surveys\footnote{http://www.lofar.org/astronomy/surveys-ksp/surveys-ksp}
(Tier-1 Wide, Tier-2 Deep, Tier-3 Ultra Deep) are expected to provide
a catalog of several hundreds candidates of galaxy clusters hosting
diffuse radio sources \citep[e.g.][]{2010A&A...509A..68C}.

The full international {\it LOFAR} telescope should be operational by
early 2011. Other new generation radio telescopes will follow in the
next few years, such as the Long Wavelength Array ({\it LWA}, 10--80
MHz)\footnote{http://lwa.unm.edu/}, the Australian Square Kilometre
Array Pathfinder ({\it ASKAP}, 70--1800
MHz)\footnote{http://www.atnf.csiro.au/SKA/}, the Karoo Array
Telescope ({\it MeerKAT}, 0.58--15
GHz)\footnote{http://www.ska.ac.za/meerkat/}. All these instruments
will indeed play an important role for the study of non-thermal
cluster physics and, more generally, will be crucial scientific and
technical pathfinders for the Square Kilometre Array ({\it SKA}, 0.10
-- 25 GHz)\footnote{http://www.skatelescope.org/}. Multi-frequency
radio surveys of the sky will be available that will unveil
statistical samples of hundreds candidates of diffuse cluster radio
sources \citep{2004NewAR..48.1137F}. As detailed above, wide and deep
complementary cluster catalogs at other wavelengths will be necessary
in order to answer the open questions about non-thermal cluster
physics.

\begin{figure*}
\resizebox{\hsize}{!}{
\includegraphics{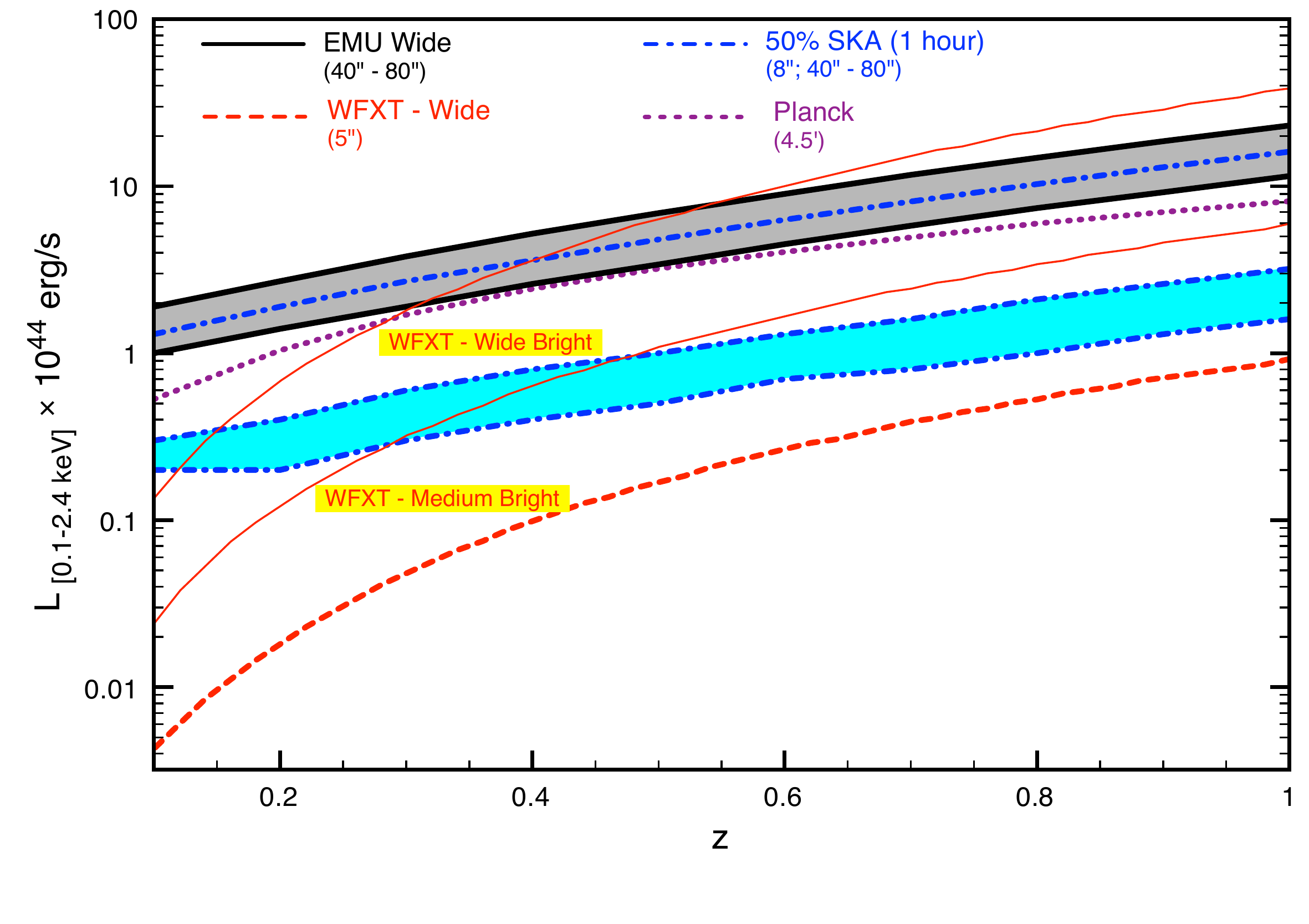}}
\caption{\footnotesize Evolution with redshift of the X-ray luminosity
  limit of clusters whose diffuse radio emission can be detected: {\bf
    a) (shaded area delimited by solid lines)} at 5-10 $\sigma$
  significance level with the {\it ASKAP} telescope down to the
  sensitivity limit of the EMU-Wide survey and at 40 to 80 arcsec
  resolution (Sect.\ref{sec:radio}); {\bf b) (dot-dahsed line)} at 10
  $\sigma$ significance level with 50\% of the {\it SKA} collecting
  area assuming an integration time of 1 hour and a resolution of 8
  arcsec \protect\citep{2004NewAR..48.1137F}; c) {\bf (shaded area
    delimited by dot-dashed lines)} at 10 $\sigma$ significance level
  with 50\% of the {\it SKA} collecting area assuming an integration
  time of 1 hour and 40 to 80 arcsec resolution. The cluster detection
  limits expected for the {\it Planck} ({\bf dotted line}; courtesy of
  A. Chamballu and J. Bartlett) and {\it WFXT}-Wide ({\bf dashed
    line}; courtesy of B. Sartoris) surveys are also shown. The
  thinner curves, finally, correspond to the bright sample limits of
  the {\it WFXT} Wide and Medium surveys
  \protect\citep{2010arXiv1003.0841S}.}
\label{fig:fig2}
\end{figure*}

After {\it LOFAR}, the following survey project very important for
non-thermal cluster studies will probably be EMU (``Evolutionary Map
of the Universe'', project leader: R. Norris). It will be a deep radio
survey ($\sim 10~\mu$Jy rms; 1130 -- 1430 MHz) covering the entire
Southern sky and part of the Northern sky ($\delta \lesssim
30^{\circ}$) with the {\it ASKAP} telescope.  Fig. \ref{fig:fig1} shows the
brightness temperature at $\approx$ 1.3 GHz as a function of redshift
expected for radio halos of a given total flux and of 0.5 Mpc
radius. The shaded area delimited by solid lines indicates an
approximate 3 $\sigma$ sensitivity level of the EMU survey. We have
taken into account that the exact observing strategy of EMU is under
discussion. The best resolution of the survey will be of $\sim$10
arcsec, but lower resolutions radio maps will also be produced in
order to increase the sensitivity to diffuse radio sources (see
Sect. 3.5.1 in \citet{2008ExA....22..151J}). We have assumed here and
in the following (see also Fig. \ref{fig:fig2}) an rms sensitivity of
10-20~$\mu$Jy/beam, with beam sizes variating from 40 to 80 arcsec.
Our estimates for the EMU survey are here compared to the T$_b \sim$
5mK sensitivity limit of 50\% of the {\it SKA} collecting area. This
3$\sigma$ sensitivity level (indicated by a dot-dashed line in
Fig. \ref{fig:fig1}) has been estimated by \citet{2004NewAR..48.1137F}
assuming an integration time of 1 hour.  The shaded area delimited by
dot-dashed curves correspond to the 3 $\sigma$ sensitivity limit of
50\% {\it SKA} at the same resolution limits that we have adopted for EMU
(from 40 to 80 arcsec).

Based on the results in Fig. \ref{fig:fig1} and on the radio halo
luminosity function derived by \citet{2002A&A...396...83E} (see also
Table 1 in \citet{2004NewAR..48.1137F}) we can expect to detect
$\gtrsim$ 300 halos at any redshift with EMU (i.e. halos in $\gtrsim
2\pi$ sterad with S$_{\rm tot} >$ 1 mJy) and several thousands
($\gtrsim$ 6000) halos with the low-resolution 50\% {\it SKA}
observations (1h integration time), among which about one third at
$z>$0.3. In such a case, in fact, our {\it SKA} estimates indicate
that we can go down to S$_{\rm tot} \approx 10 \mu$Jy at any
redshift. Note that, in addition, EMU could detect several tens higher
redshift ($\gtrsim$ 0.3) halos with S$_{\rm tot} >$ 0.1 mJy.

We have then refined our estimates to evaluate the evolution with
redshift of the X-ray luminosity limit of clusters whose diffuse radio
emission can be detected by the EMU Survey (shaded are delimited by
solid curves in Fig.~\ref{fig:fig2}).  We have considered radio halos
of 1 Mpc size with radio luminosities ${\rm L}_{1.4 {\rm GHz}} \gtrsim
5 \times 10^{20}$ W/Hz and a typical brightness profile as a function
of radius has been adopted \citep{2001A&A...369..441G}:

$$
B_{\nu} (\eta {\rm R}_h) = \xi \frac{{\rm L}_{1.4 GHz} (1400/\nu)^{\alpha} (1+z)^{-(3+\alpha)}}{1.5 \times 10^{31} {(\eta {\rm R}_h )}^2}
$$

\noindent where $\xi$ indicates the fraction of the total flux of the
source at $r=\eta {\rm R}_h$ (thus $\xi\leq1$ and $\eta\leq 1$),
B$_{\nu}$ is in Jy/arcsec$^2$, $L_{1.4 GHz}$ in W/Hz, $\nu$ in MHz and
R$_h$ (=0.5) in Mpc. In our estimates a radio halo is considered to be
detected when $B_{\nu} (\eta {\rm R}_h) \geq$ 5-10 rms$_{\rm EMU}$ and
$\xi=0.5$.  The EMU detection limits for radio halo luminosities have
finally be converted to X-ray luminosities of the host clusters
following Eq. (1) in \citet{2006MNRAS.369.1577C}. Increased inverse
Compton energy losses on the CMB at higher redshift and the consequent
decrease in the intrinsic radio halo luminosity have also been taken
into account \citep{2002A&A...396...83E}.

Note that the EMU limits shown in Fig.~\ref{fig:fig2} concern the Wide
EMU survey described above. The possibility to perform deeper {\it
  ASKAP} surveys, in particular at lower frequencies ($\sim$850 MHz)
that are more favorable to radio halo detection, is currently
considered within the EMU project (Johnston-Hollitt, private
communication). In such a case, complementary cluster catalogs could
provide excellent targets for deeper radio follow-ups, thus helping in
selecting the regions of the sky to be observed with {\it ASKAP}, {\it
  MeerKAT} or any other radio facility.  Fig. \ref{fig:fig2} also
shows the detection limit expected with 50\% of the {\it SKA}
collecting area at a resolution of $\sim$8, 40 and 80 arcsec, assuming
a bandwidth of 0.5 GHz and an integration time of 1 hour. As before, a
3$\sigma$ level brightness sensitivity of $T_b \sim$ 5 mK has been
adopted for these estimates \citep{2004NewAR..48.1137F}. We have
converted this radio brightness sensitivity to X-ray luminosity limits
as a function of redshift by adopting exactly the same method of our
previous EMU estimates. The detection limits are here at 10 $\sigma$
level. Much deeper sensitivity limits can of course be reached with
the full {\it SKA} and longer integration times.

\subsection{Importance of {\it WFXT} surveys} \label{sec:wfxt}

In order to be able to test current models about the origin of the
non-thermal intracluster component we need both {\it statistical studies} of
the fraction of observed clusters hosting diffuse radio sources as a
function of the cluster mass and $z$, and {\it detailed analyses} of how
the correlation between the radio emission and the physical properties
of clusters ($L_X$, $T_X$, mass, dynamical state...) evolve with $z$.
The newly identified candidates of radio emitting clusters described
in Sect. \ref{sec:radio} will thus have to be cross-matched with cluster
catalogs in other wave bands, which will be needed for the cluster
{\it identification} and {\it physical characterization}.

On short timescales, existing or incoming optical, IR, sub-mm and
X-ray surveys will provide an important set of complementary data for
the identification of potential clusters detected by {\it LOFAR}
\citep[among
others:][]{1998MNRAS.301..881E,2000ApJS..129..435B,2001ApJ...553..668E,2001A&A...369..826B,2002AJ....123.1807G,2006MNRAS.373L..26V,2007A&A...461...81O}.
In \citet{2010arXiv1005.3699F} we have compared the cluster X-ray
luminosity detection limits of {\it Planck} and {\it LOFAR}.  Our
estimates indicate that the Tier-1 Wide {\it LOFAR} survey will
provide a galaxy cluster catalog through diffuse radio source
detection that well match the expected {\it Planck} cluster detection
at $z\lesssim0.3$. At higher redshift, all the systems detected by
{\it LOFAR} will have an X-ray luminosity above the {\it Planck}
detection limit.  Fig. \ref{fig:fig2} shows that the cluster detection
limits of {\it Planck} will also be perfectly suited for the
comparison with the list of diffuse radio surveys resulting from the
EMU Wide survey. {\it
  eROSITA}\footnote{http://www.mpe.mpg.de/heg/www/Projects/\\EROSITA/main.html}
detection limits should match well the radio halo cluster detection
limits with EMU presented here (Reiprich, private
communication). Based on our estimates radio observations with the SKA
will instead require deeper complementary surveys for cluster
cross-identification.

The projected {\it WFXT} surveys will provide two kinds of cluster
catalogs \citep{2009astro2010S..90G,2010arXiv1003.0841S}: $\sim 3
\times 10^6$ detected clusters (out of which $\approx$ 98\% at
$z<1$\footnote{For $z \gtrsim 1$ the lifetime of \cres~ whose
  synchrotron emission peaks at 1.4 GHz $\tau \lesssim$ 10 Myr due to
  IC energy losses.}), and $\sim 2 \times 10^4$ clusters (the so
called ``bright sample''), which, with a flux limit 30 times brighter
than the detection flux limit, will have robust measures of mass
proxies, as well as of ICM surface brightness and temperature
profiles.

Most of the cluster detections will come from the all-sky {\it WFXT}
Wide Survey, while the ``bright sample'' at $z<1$ is mainly due to the
Medium {\it WFXT} survey, which would cover 3000 square degrees (see
Table 1 and Fig. 3 in \citet{2010arXiv1003.0841S}). In
Fig. \ref{fig:fig2} we have plotted the X-ray luminosity detection
limits of the {\it WFXT} Wide survey as a function of $z$ (red dashed
curve), as well as the X-ray luminosity limits for the bright samples
resulting from the Wide and Medium {\it WFXT} surveys (thin red
curves).  Fig. \ref{fig:fig2} shows that the comparison between
possible {\it WFXT} and radio surveys could provide:

\begin{itemize}

\item an all-sky cluster catalog ({\it WFXT} Wide, dashed thick line in
  Fig. \ref{fig:fig2}) deep enough for the identification of
  $\gtrsim$6000 candidate clusters hosting diffuse radio emission
  coming from {\it SKA} observations (see Sect. \ref{sec:est} and
  Fig. \ref{fig:fig1}). This sample will offer the unique opportunity
  to study in a fully statistical way the cluster radio {\it vs.}
  X-ray luminosity correlation (Sect. \ref{sec:open});

\item the possibility to compare radial profiles of the radio spectral
  index $\alpha$ and of the ICM brightness and temperature
  (Sect. \ref{sec:open}). This could be done on several hundred
  clusters at $z<$0.5 by combinining radio surveys data ({\it LOFAR},
  {\it ASKAP}, {\it SKA} in Fig. \ref{fig:fig2}) with the bright
  samples deriving from {\it WFXT} Wide and Medium surveys (thin red
  lines in Fig. \ref{fig:fig2});

\item interesting targets for deeper {\it LOFAR}, {\it ASKAP}, {\it
    MeerKAT} or {\it SKA} follow-ups.

\end{itemize}

\noindent WFXT will provide X-ray surveys with the necessary
sensitivity to match those achievable in future radio surveys of
galaxy clusters.

\begin{acknowledgements}
  I am very grateful to M. Arnaud, S. Borgani, G.  Giovannini,
  M. Johnston-Hollitt and the EMU cluster working group, P. Rosati and
  B. Sartoris for very useful discussions that helped to improve the
  paper. I acknowledge financial support by the Agence Nationale de la
  Recherche through grant ANR-09-JCJC-0001-01.
\end{acknowledgements}

\bibliographystyle{aa}

\end{document}